\documentclass[showpacs, pra,twocolumn,preprintnumbers ,amsmath, amssymb, superscriptaddress, aps]{revtex4-2}
\usepackage{color}
\usepackage{amsmath,amssymb}
\usepackage{pifont}
\usepackage{amssymb}  
\usepackage{bbold}
\usepackage{float}
\usepackage{subfloat}
\usepackage[squaren]{SIunits}
\usepackage[colorlinks=true,linkcolor=blue,citecolor=blue]{hyperref}%
\usepackage[caption=false]{subfig}
\usepackage{tikz}
\usepackage{makecell}
\usepackage{subfig}
\usepackage{pifont}   
\usepackage{graphicx} 
\graphicspath{{Figures/}}
\usepackage{dcolumn}  
\usepackage{bm}       
\usepackage{multirow} 
\usepackage{placeins}
\usepackage{mathtools}

\DeclareUnicodeCharacter{0001}{\textbf{}}
\begin{document}	
\title{
	Transport properties of hybrid single-bilayer graphene interfaces in magnetic field}
	\date{\today}
	\author{Nadia Benlakhouy}
	\email{benlakhouy.n@ucd.ac.ma}
	\affiliation{Laboratory of Theoretical Physics, Faculty of Sciences, Choua\"ib Doukkali University, PO Box 20, 24000 El Jadida, Morocco}
	\author{Ahmed Jellal}
	\email{a.jellal@ucd.ac.ma}
	\affiliation{Laboratory of Theoretical Physics, Faculty of Sciences, Choua\"ib Doukkali University, PO Box 20, 24000 El Jadida, Morocco}
	\affiliation{Canadian Quantum  Research Center,
				204-3002 32 Ave Vernon,  BC V1T 2L7,  Canada}
			
				\author{Michael Schreiber}
			\affiliation{Institut für Physik, Technische Universität, D-09107 Chemnitz, Germany}

\pacs{73.22.Pr, 72.80.Vp, 73.63.-b\\
{\sc Keywords}: Graphene junctions, magnetic field, energy spectrum, transmission, conductance, Klein tunneling.}
	\begin{abstract}
		We investigate the electronic properties of a hybrid system that comprises single-bilayer graphene structures subjected to a perpendicular magnetic field. Specifically, our focus is on the behavior exhibited by the zigzag boundaries of the junction, namely Zigzag-1 (ZZ1) and Zigzag-2 (ZZ2), using the continuum Dirac model for rigorous analysis. Our findings reveal a striking dependence of conductance on the width of the bilayer graphene at ZZ1, providing essential insights into the transport behavior of this boundary. Moreover, we observe a captivating phenomenon where the conductance at ZZ2 exhibits prominent maxima, demonstrating a robust correlation with the applied magnetic field. Additionally, our investigation uncovers the profound impact of interfaces on transmission probability, with ZZ1 being notably more affected compared to ZZ2. The variation of the Fermi energy further highlights the significant influence of magnetic field strength on the system's conductive properties, resulting in distinct conductance characteristics between the two regions. The combined results of ZZ1 and ZZ2 provide valuable insights into the system's transport properties. Notably, a clear exponential-like trend in conductance variation with the applied magnetic field underscores the system's strong sensitivity to magnetic changes.
	\end{abstract}
	\maketitle
\section{Introduction}
Graphene is a two-dimensional, single-layer sheet of carbon atoms arranged in a hexagonal lattice structure \cite{A. K. Geim}. It is the thinnest, strongest, and most conductive material known to science, with exceptional mechanical, electrical, and thermal properties. Graphene has sparked widespread advancements in various disciplines due to  its potential applications \cite{A. K. Geim, K.S. Novoselov, R.R. Nair, M.I. Katsnelson, O. Klein}. Bilayer graphene (BLG) is a material composed of two layers of graphene sheets stacked on top of each other. The two layers are separated by a small {interlayer hopping $\gamma_{1}$}, and can exhibit different electronic properties depending on the stacking order. When the two graphene layers are aligned in the same direction, with their atoms directly on top of each other, it is called AA-BLG. This results in perfect lattice symmetry, which gives rise to a special electronic band structure that depends on the interlayer spacing \cite{J.-K. Lee, J. Borysiuk, Z. Liu}. Another interesting stacking is AB-BLG. In the AB-BLG configuration, the atoms in the two layers are not aligned with each other, leading to a slight variation in the electronic properties of the bilayer compared to single-layer graphene (SLG).
{Indeed, in the absence of an electric field, the electronic structure of the material remains gapless at the K and K' points. The opening of a band gap occurs when an external electric field is applied, allowing for tunability of the electronic properties \cite{W. Zhang}. Similar results hold for large twist angles \cite{T. Spenser}}.
The development of high-quality samples and theoretical and experimental research into its distinctive electronic properties make AB-BLG appealing for a variety of applications \cite{T. Ohta, M. O. Goerbig, H. M. Abdullah, M. Hassane Saley, E. McCann, E. McCann1,  A. Rozhkov, I. Redouani, N. Benlakhouy}.
{
Another intriguing area of research is on twisted bilayer graphene (TBG), where the two graphene layers are rotated at a specific twist angle relative to each other. This twist-induced moiré pattern results in a tunable electronic band structure, leading to the emergence of novel electronic states such as Mott insulators, superconductors, and topological phases \cite{A.Nimbalkar}. Additionally, the study of transition metal dichalcogenides (TMDs) has gained significant attention in recent years. TMDs are a class of two-dimensional materials with a structure similar to graphene, but they consist of transition metal atoms sandwiched between two layers of chalcogen atoms \cite{S. Manzeli}.}

Recent studies have shown that junctions between regions of different numbers of graphene layers, such as the SLG/BLG interfaces,  can result in interesting properties.  For instance, in \cite{Nakanishi} the transmission probability through SLG/BLG junction was estimated in the absence of a magnetic field. Theoretical investigations on the transport characteristics of BLG with locally decoupled graphene sheets have also been performed  \cite{H. M. Abdullah}. Moreover, a BLG flake sandwiched between two single zigzag or armchair nanoribbons was studied, and it was shown that oscillations in the conductance were seen at energies greater than the interlayer coupling \cite{J. W. Gonzalez}. Recently, it was found that the interface of these hybrid systems exhibits an unconventional Landau quantization \cite{W. Yan, L.-J. Yin}. Another study was devoted to  the effects of an electric bias and a perpendicular magnetic field on the electron energy spectrum in SLG/BLG and BLG/SLG structures \cite{M. Mirzakhani}. It is generally known that graphene has two main types of edges: zigzag and armchair edges. Regarding the electrical structure of finite-sized systems, it has been established that graphene with zigzag edges exhibits  localized states close to  Fermi energy, but those with armchair edges do not \cite{S.E. Stein, K. Tanaka, M. Fujita,  M. Fujita1, K. Nakada, H.-X. Zheng, N.M.R. Peres,  M. Kohmoto, S. Ryu, K. Sasaki, K. Sasaki1, A.H. Castro Neto, Y. Niimi, Y. Kobayashi, Z-x. Hu}. As a  consequence, the existence of an edge state results in notable variations in the transport characteristics.

Various works on the edge states of the hybrid interface attracted our interest. Inspired by the findings in Refs. \cite{M. Mirzakhani}, and \cite{Nakanishi}, we investigate the transport properties of SLG and AB-BLG junctions that can be created from the block shown in Fig. \ref{SLG-BLG-SLG}. {Our study is centered on a unique configuration, where SLG interfaces with an AB-BLG segment subjected to a magnetic field.
It is imperative to note that our configuration highlights
a notably asymmetric distribution of the magnetic field.
Specifically, the magnetic field exists solely on AB-BLG
while being absent in SLG. This asymmetric distribution
of the magnetic field within the model stands as a distinct
feature, influencing the transport properties under examination \cite{M. Ramezani Masir}.}
{Note that a region with a sharply cut-off magnetic field is difficult to realize experimentally. First, we would like to point out that, due to the analytical nature of our work, we had to use some simplifications to work on a problem that is analytically tractable. This, of course, is only useful if the simplifications that were used in a way mimic an experimentally possible problem. In our case, this would mean a problem with a magnetic field in the bilayer region and no magnetic field in the single-layer graphene region. Such a setup can be achieved in an approximate way as follows: one could imagine putting a superconductor as a shield above the single-layer regions. If the superconductor is thick enough, no magnetic field, or at least very little, could penetrate according to the London penetration depth. Indeed, one would also have to expect not fully sharp edges of the field near the transitions between bilayer and single-layer graphene regions. However, such complicated position-dependent fields would be difficult to treat analytically. Therefore, we chose to come up with a simplified setup that mimics the shape one would expect to a good extent—a sharp drop in magnetic field. Of course, this is not fully accurate, but for the purposes of an analytical model that captures the main features of our idea, it should be fine. Indeed, similar idealizations are made in any standard quantum mechanics course when one deals with the tunneling problem. No potential will be realizable that has perfectly sharp edges. Nevertheless, in many situations, it can be a relatively good approximation.
}

{As seen in Fig. \ref{SLG-BLG-SLG}, the structure's ZZ junctions} cannot have the same edge interface on both sides, therefore, they always have a pair of distinct ZZ boundaries, denoted ZZ1 and ZZ2. When considering this distinction, {within the ZZ1 border, our conductance demonstrates a dependency on energy and exhibits antiresonances, approaching nearly zero under high magnetic field},
because of the coexistence of two propagating channels. {Our investigation demonstrates an intriguing conductance behavior in the bilayer graphene (BLG) system. This is due to the influence of both the BLG width and the magnetic field that is being used. Specifically, we have observed that the conductance exhibits a notable dependence on the width of the BLG, particularly at higher energies. This phenomenon arises due to the quantum confinement effect, which modifies the electronic states and energy levels within the system. Furthermore, the introduction of a magnetic field introduces a fascinating aspect to conductance behavior. The resonances in the transmission measurements become significant and display a clear dependence on the strength of the magnetic field.}
 Turning now to the ZZ2 boundary, the conductance as a function of the magnetic field shows maxima, in contrast to the ZZ1 boundary, and by increasing the width of BLG and the Fermi energy, the conductance shows oscillation in the ZZ2 feature.  the transmission probability is substantially influenced by boundaries and the ZZ1 boundary's confinement is more significant than the ZZ2 boundary's. {Our analysis of the conductance behavior for ZZ1 and ZZ2, varying with the Fermi energy, revealed interesting trends. Notably, ZZ1 demonstrated higher conductance compared to ZZ2, while both regions showed a rapid decrease in conductance with increasing Fermi energy. These findings highlight the significant impact of the magnetic field on the conductive properties of ZZ1 and ZZ2, resulting in distinct conductance characteristics between the two regions. Our combined results of ZZ1 and ZZ2 presented in this study offer valuable insights into the overall transport properties of the system under investigation. One of the key observations is the conductance variation with respect to the applied magnetic field. Notably, a clear and consistent exponential-like trend emerges, revealing the system's strong sensitivity to changes in the magnetic field.}
  
 {Our findings may shed light on the intriguing phenomenon of Fabry-Perot oscillations in transmission behavior, which emerge as a consequence of quantum interference effects. Significantly, our work unveils a critical threshold dictated by the interlayer hopping term, below which a single propagating mode dominates, leading to the absence of oscillations or interference. The comprehensive analysis of the length, energy, and magnetic field influences on the transmission provides valuable insights into the intricate behaviors of this hybrid graphene system. These novel findings hold immense implications for various applications and research areas within the scientific community.}

The present paper is structured as follows. In Sec.  \ref{THEORY AND MODEL}, we consider an SLG/AB-BLG/SLG structure and use the full-band continuum model to establish the energy spectrum. We introduce formulation to describe two kinds of zigzag boundaries of the junction, zigzag-1 (ZZ1) and zigzag-2 (ZZ2)  in Sec. \ref{Single-layer and bilayer junction}. Sec. \ref{RESULTS AND DISCUSSION} is devoted to the numerical analysis of our findings and comparison with literature. 
In Sec. \ref{Conclusion}, we summarize our main conclusions. In Appendix,  
we mainly formulate the details of the transfer matrix method for the SLG and BLG interfaces.
\section{THEORY AND MODEL}\label{THEORY AND MODEL}
\subsection{Bilayer graphene}
\begin{figure*}
	\centering
	\subfloat[]{\includegraphics[width=0.4\linewidth]{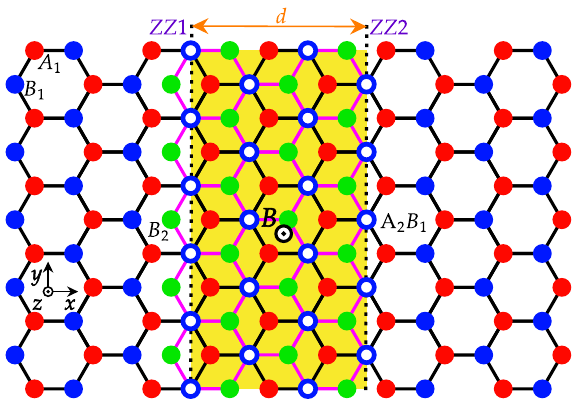}}\ \ \ \
	\subfloat[]{\includegraphics[width=0.4\linewidth]{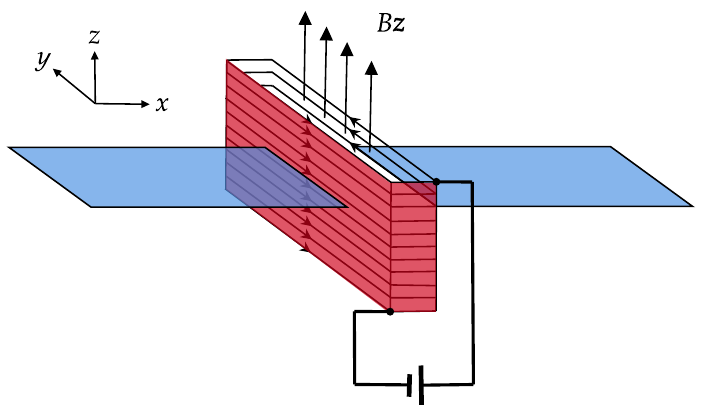}}
	{\caption{(Color online)  (a) Schematic presentation of hybrid SLG/BLG interfaces of types ZZ1 and ZZ2, consisting of  SLG connected to an AB-BLG subjected to a magnetic field $B$. The BLG width is $d$ (yellow region). Black thick lines represent the bottom layer with $A_1$ (red) and $B_1$ (blue) sites, whereas pink lines represent the top layer with $A_2$ (white) and $B_2$ (green) sites. (b)  Diagram illustrating a configuration in which a coil with a rectangular cross-sectional shape and a very narrow width generates a magnetic field that closely resembles the one in our specific scenario. 
		}\label{SLG-BLG-SLG}}
\end{figure*}
AB-BLG is a type of bilayer graphene in which the two layers are stacked in an AB arrangement. In this stacking configuration, the carbon atoms in one layer sit directly above the centers of the hexagons in the other layer, resulting in a characteristic Bernal stacking pattern \cite{J.D. Bernal}. It contains $A_1$ and $B_1$ atoms on layer $1$ and $A_2$ and $B_2$ on layer $2$, which are connected by interlayer coupling $\gamma_{1}$. 

To achieve our goal, we consider the geometry depicted in Fig. \ref{SLG-BLG-SLG}. Then, without taking into account the minor contributions of the other interlayer couplings, the effective Hamiltonian  in the vicinity of the $K$ valley is given by \cite{S. Sarma, E. J. Mele}
\begin{equation}\label{1}
\mathcal{H}_{\text{BLG}}=\begin{pmatrix}
0 & v_{F}\pi^{\dag} & 0 & 0 \\
v_{F}\pi & 0 & \gamma_{1} & 0\\
0 & \gamma_{1} & 0 & v_{F}\pi^{\dag} \\
0 & 0 &  v_{F}\pi& 0 \\
\end{pmatrix}
\end{equation}
where {the momentum $\pi=k_x+ik_y$ ($\hbar =1$)},  $v_{F}=10^{6}$ m/s is the Fermi velocity for electrons in each graphene layer, and $\gamma_1=0.4$ eV is the nearest-neighbor interlayer hopping term. The eigenstates of  $\mathcal{H}_{\text{BLG}}$ are four-component spinors $\Psi(x,y)=\left[\psi_{A_{1}},\psi_{B_{1}},\psi_{A_{2}},\psi_{B_{2}} \right]^{T}$. 
In the presence of  a constant  magnetic field which is described by 
\begin{equation}
B(x)= B\Theta(x-d)
\end{equation}
 one  substitutes the canonical momentum $\mathbf{p}$ by the gauge-invariant kinetic momentum $\mathbf{p} + e\mathbf{A}$ in Eq. \ref{1}. $\mathbf{A}=(0, Bx)$ is the vector potential chosen in the Landau gauge.
We may solve the eigenvalue problem by separating the variables and writing the eigenspinors as a plane wave in the $y$-direction because of the conservation of $p_y$. As a result, we write
$
\Psi(x, y)=e^{i k_y y} \psi(x, k_y)
$ and
 the envelope functions $\psi(x, k_y) \equiv \psi(X)$ depend only on a single combination of the variables, $X=\frac{x}{\ell_B}+k_y \ell_B$, with 
$\ell_B=\sqrt{1 /(e B)}$ being the magnetic length. They satisfy the eigenvalue equation
\begin{equation}\label{2}
\left(\begin{array}{cccc}
0 & \sqrt{2}\epsilon_0 \hat{a} & 0 & 0 \\
\sqrt{2}\epsilon_0 \hat{a}^{\dagger} & 0 & \gamma_1 & 0 \\
0 &  \gamma_1 & 0 &  \sqrt{2}\epsilon_0 \hat{a} \\
0 & 0 & \sqrt{2}\epsilon_0 \hat{a}^{\dagger} & 0
\end{array}\right) \Psi=E \Psi,
\end{equation}
where the annihilation $\hat{a}=\frac{1}{\sqrt{2}}\left(X+\partial_X\right)$ and creation $\hat{a}^{\dagger}=\frac{1}{\sqrt{2}}\left(-\partial_X +X\right)$  operators are fulfilling the commutation relation $[\hat{a}, \hat{a}^{\dagger}]=\mathbb{1}$, with  $\epsilon_0= v_F /\ell_B$.
Now from Eq. \ref{2}, we obtain four coupled equations  as
\begin{align}
\label{31} -i\sqrt{2}\epsilon_0 \hat{a} \psi_{B_1}(X) &=E \psi_{A_1}(X),\\ 
\label{32}i\sqrt{2}\epsilon_0 \hat{a}^{\dagger}\psi_{A_1}(X)+\gamma_1 \psi_{A_2}(X)&=E \psi_{B_1}(X),\\
-i\sqrt{2}\epsilon_0 \hat{a} \psi_{B_2}(X)+\gamma_1 \psi_{B_1}(X) &=E \psi_{A_2}(X), \\
\label{34}i\sqrt{2}\epsilon_0 \hat{a}^{\dagger} \psi_{A_2}(X) &=E \psi_{B_2}(X).
\end{align}
By eliminating $\psi_{A_1}(X), \psi_{A_2}(X)$, and $\psi_{B_2}(X)$, we get  the fourth order differential equation
\begin{equation}
\begin{array}{r}
{\left[2\epsilon_0^2 \hat{a}^{\dagger}\hat{a}-E^2\right]\left[2\epsilon_0^2 \hat{a} \hat{a}^{\dagger}-E^2\right] \psi_{B_1}}(X) 
=\gamma_1E^2 \psi_{B_1}(X),
\end{array}
\end{equation}
or equivalently
\begin{equation}\label{4}
\left(\partial_X^2-X^2-1-2\lambda_{+}\right)\left(\partial_X^2-X^2-1-2 \lambda_{-}\right)\psi_{B_1}(X)=0,
\end{equation}
where $\lambda_{\pm}$ defines the energy bands
\begin{equation}
\lambda_{\pm}= -\frac{1}{2}+\frac{E^2}{2\epsilon_0^2} 
\pm \frac{\sqrt{\epsilon_0^4+ \gamma_1^2E^2}}{2 \epsilon_0^2},
\end{equation}
{Therefore, by solving Eq. \ref{4}, we can obtain the energy spectrum
\begin{equation} \label{5}
E_{n}^{ \pm}= \pm\left\{\epsilon_0^2(2n+1)+\frac{\gamma_1^{ 2}}{2} \pm\sqrt{\frac{\gamma_1^{4}}{4}+(2n+1) \gamma_1^{ 2}\epsilon_0^2+\epsilon_0^4}\right\}^{1/2}
\end{equation}
where $n$ is an integer, the Landau level index.  For $\gamma_1\rightarrow0$, the last equation reduces to that of single layer with spectrum 
\begin{equation}
E=\pm\epsilon_0\sqrt{2n+1\pm1}.
\end{equation}
Additionally,} the general solution of Eq. \ref{4} can be expressed in terms of Weber’s parabolic cylinder function $D_\lambda(Z)$ \cite{DLMF}, where $Z=\sqrt{2}X$.
Therefore, by solving the Eq. \ref{4} we can  obtain the energy as done by \cite{I. Redouani} with considering just the magnetic field, and also the general solution of Eq. \ref{4} can be  written in terms of Weber’s parabolic cylinder function $D_\lambda(Z)$ \cite{DLMF}, with $Z=\sqrt{2} X$.
Thus, we have
\begin{equation}\label{Eq10}
\psi_{B_1}(Z)=\psi_{B_1}^{+}(Z)+\psi_{B_1}^{-}(Z),
\end{equation}
with
\begin{equation}
\begin{aligned}
&\psi_{B_1}^{+}(Z)=c_{+} D_{\lambda_{+}}(Z)+c_{-} D_{\lambda_{+}}(-Z), \\
&\psi_{B_1}^{-}(Z)=d_{+} D_{\lambda_{-}}(Z)+d_{-}D_{\lambda_{-}}(-Z),
\end{aligned}
\end{equation}
with the constants $c_\pm$ and $d_\pm$.
The rest of the  components can derived using the coupled equations.  $\psi_{A_1}(Z)= \psi_{A_1}^{+}(Z)+ \psi_{A_1}^{-}(Z)$ can be obtained by substituting  Eq. \eqref{Eq10} into Eq. \ref{31}
\begin{equation}
\begin{aligned}
&\psi_{A_1}^{+}(Z)=c_{+} \nu \lambda_{+} D_{\lambda_{+}-1}(Z)+c_{-} \nu^* \lambda_{+} D_{\lambda_{+}-1}(-Z),\\
&\psi_{A_1}^{-}(Z)=d_{+} \nu\lambda_{-}   D_{\lambda_{-}-1}(Z)+d_{-}  \nu^* \lambda_{-} D_{\lambda_{-}-1}(-Z),
\end{aligned}
\end{equation}
where $ \nu=-\frac{i \sqrt{2} \epsilon_0}{E}$. Combining  $\psi_{A_1}$ and $\psi_{B_1}$ in Eq. \ref{32} gives $\psi_{A_2}(Z)= \psi_{A_2}^{+}(Z)+\psi_{A_2}^{-}(Z)$
with
\begin{equation}
\begin{aligned}
&\psi_{A_2}^{+}(Z)=c_{+} \zeta^{+} D_{\lambda_{+}}(Z)+c_{-} \zeta^{+}  D_{\lambda_{+}}(-Z), \\
&\psi_{A_2}^{-}(Z)=d_{+} \zeta^{-}  D_{\lambda_{-}}(Z)+d_{-} \zeta^{-}  D_{\lambda_{-}}(-Z),
\end{aligned}
\end{equation}
and $\zeta^{\pm}=\frac{E}{\gamma_1}-\frac{2 \epsilon_0^2 \lambda_{\pm}}{\gamma_1 E}$. Finally from Eq. \ref{34}, we obtain 
$
\psi_{B_2}(Z)=\psi_{B_2}^{+}(Z)+ \psi_{B_2}^{-}(Z),
$
with
\begin{equation}
\begin{aligned}
&\psi_{B_2}^{+}(Z)=c_{+} \nu^{*} \zeta^{+}  D_{\lambda_{+}+1}(Z)+c_{-} \nu^{*}\zeta^{+}  D_{\lambda_{+}+1}(-Z), \\
&\psi_{B_2}^{-}(Z)=d_{+} \nu \zeta^{-} D_{\lambda_{-}+1}(Z)+d_{-} \nu \zeta^{-}\lambda_{-}  D_{\lambda_{-}+1}(-Z).
\end{aligned}
\end{equation}
\subsection{Single-layer graphene}
\begin{figure}
	\centering
\subfloat[]{\includegraphics[width=0.32\linewidth]{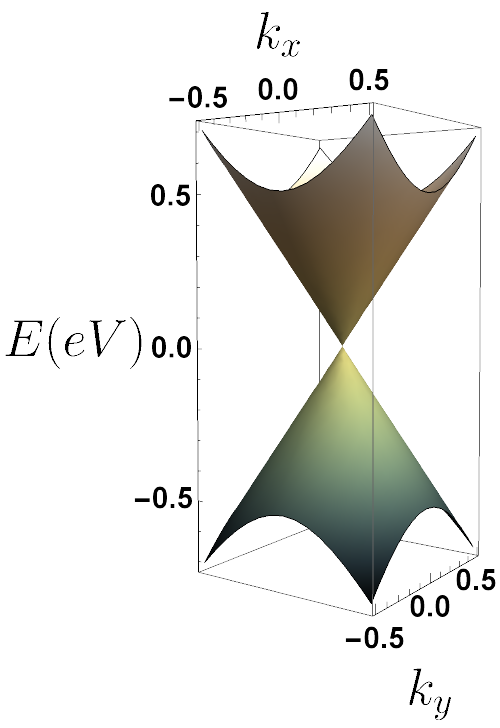}}
\subfloat[]{\includegraphics[width=0.62\linewidth]{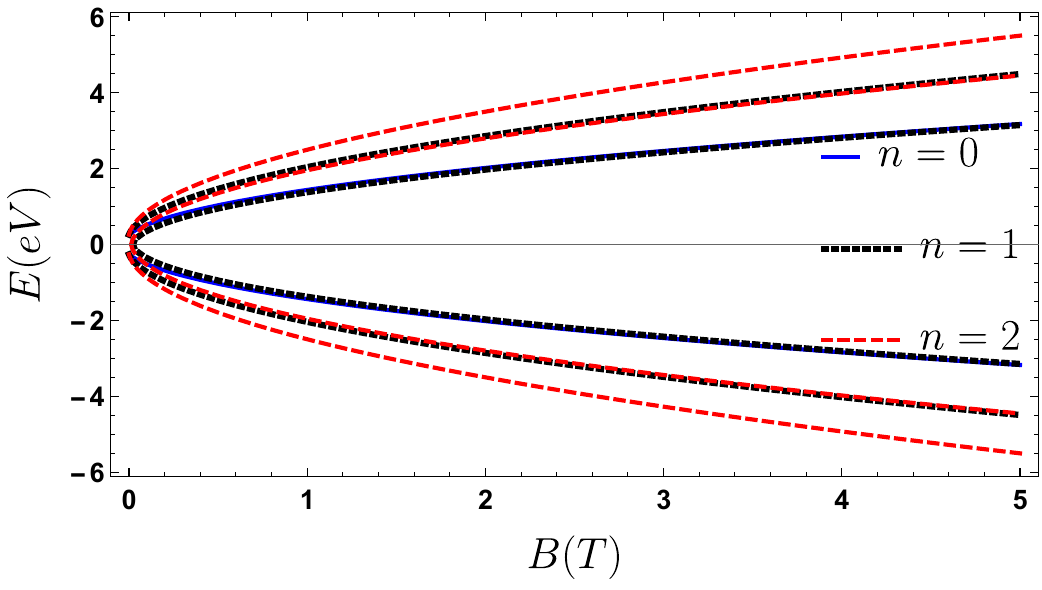}}
{\caption{(Color online)  Spectrum representations:  (a) SLG as function of wave vector  and 
		(b)  BLG as a function of magnetic field.
	}\label{spectrum}}
	\end{figure}

 The eigenspinors  of SLG  are given by solving the time-independent Schr\"odinger equation for the Hamiltonian 
\begin{equation}
{\mathcal{H}_{\text{SLG}}=v_F \boldsymbol{k}\cdot\boldsymbol{\sigma},} \end{equation}
to end up with the wave function
\begin{equation}
\Phi(x, y)=e^{ik_yy}\begin{pmatrix}
\phi_A(x) \\
\phi_B(x)
\end{pmatrix},
\end{equation}
where the two  $x$-dependent components are given by
\begin{align}\label{phiA-PhiB}
&\phi_A(x)=a\alpha^{-}e^{ik_xx}-b\alpha^{+}e^{-ik_xx},\\
&\phi_B(x)=ae^{ik_xx}+be^{-ik_xx},	
\end{align}
and we have 
\begin{equation}
k_{x}=\sqrt{E^{2}-k_{y}^{2}},\quad \alpha^{\pm}=\frac{k_{x}\pm ik_{y}}{E}
\end{equation}
with two constants $a$ and $b$. {Fig. \ref{spectrum} provides a depiction of the energy spectrum for both systems. Specifically, Fig. \ref{spectrum} (a) illustrates the dispersion relation for SLG as a function of wave vector, while Fig. \ref{spectrum} (b) displays the first three lowest levels of BLG in relation to the magnetic field $B$. It is worth nothing that a comparable result for BLG can be found in \cite{Redouani2}.}

\section{SLG and BLG junction}\label{Single-layer and bilayer junction}
Our presumption is that charge carriers always move from left to right. We consider a system that combines  SLG and  BLG. In this system, the leads on the left and right are SLGs, while in between they are connected to an AB-BLG subjected to a magnetic field. In the following, we take into consideration two different zigzag boundary types: zigzag-1 (ZZ1) and zigzag-2 (ZZ2).
\subsection{Zigzag boundary, ZZ1}
{The front-most line of the bilayer edge is created by $B_1$ and $A_2$ sites, as shown in Fig. \ref{SLG-BLG-SLG}}. Note that ZZ1 is aligned to the honeycomb lattice's zigzag direction. Then we use the continuity of wavefunctions to obtain at $x=0$
\begin{align}\label{ZZ1x0}
	& \phi_{\text{A}}(x=0)=\psi_{\text{A}_1}(Z_{1}), \\
	&\phi_{\text{B}}(x=0)=\psi_{\text{B}_1}(Z_{1}),\\
	&\psi_{\text{B}_2}(Z_{1})=0,
\end{align}
and at $x=d$
\begin{align}\label{ZZ1xd}
	&\psi_{\text{A}_{1}}(Z_{2})=\phi_{\text{A}}(x=d), \\
	&\psi_{\text{B}_{1}}(Z_{2})=\phi_{\text{B}}(x=d),\\
	&\psi_{\text{B}_2}(Z_{2})=0,
\end{align}
where we have set $Z_{1}=\sqrt{2}k_{y}\ell_{B}$ and
$Z_{2}=\sqrt{2}(\frac{d}{\ell_B}+k_y \ell_B)$.
\subsection{Zigzag boundary, ZZ2}
{As far as  ZZ2 is concerned, $B_2$ sites form the front-most line of the bilayer region (Fig. \ref{SLG-BLG-SLG})}. Also, the continuity leads to set of equations \cite{Nakanishi, Koshino}
\begin{align}\label{ZZ2x0}
	& \phi_{\text{A}}(x=0)=\psi_{\text{A}_1}(Z_{1}), \\
	&\phi_{\text{B}}(x=0)=\psi_{\text{B}_1}(Z_{1}),\\
	&\psi_{\text{A}_2}(Z_{1})=0,
	\\
	\label{ZZ2xd}
	&\psi_{\text{A}_{1}}(Z_{2})=\phi_{\text{A}}(x=d), \\
	&\psi_{\text{B}_{1}}(Z_{2})=\phi_{\text{B}}(x=d),\\
	&\psi_{\text{A}_2}(Z_{2})=0.
\end{align}

The above matching equations are worked out in the Appendix  to
to establish  the transmission probability $T(E)$ for each boundary. Then a transfer matrix approach was used to get the two transmission coefficients given in Eqs. \eqref{A18} and \eqref{A29}. They can serve to derive 
the conductance based on the Landauer-Buttiker formula
\begin{equation}
	G(E)=G_{0}T(E),
\end{equation}
with the unit $G_{0}=2e^{2}/h$.
\begin{figure}[htp]
	\centering
	\includegraphics[width=0.5\linewidth]{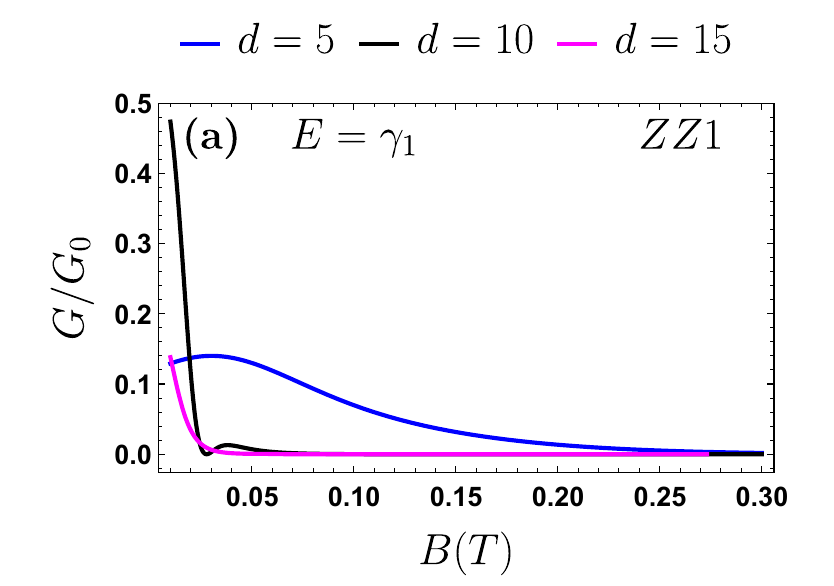}\includegraphics[width=0.5\linewidth]{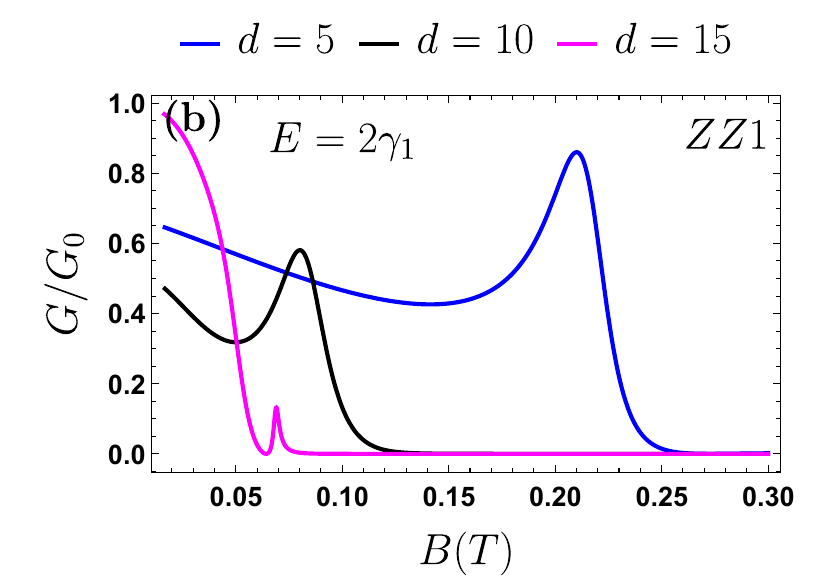}\\
	\includegraphics[width=0.5\linewidth]{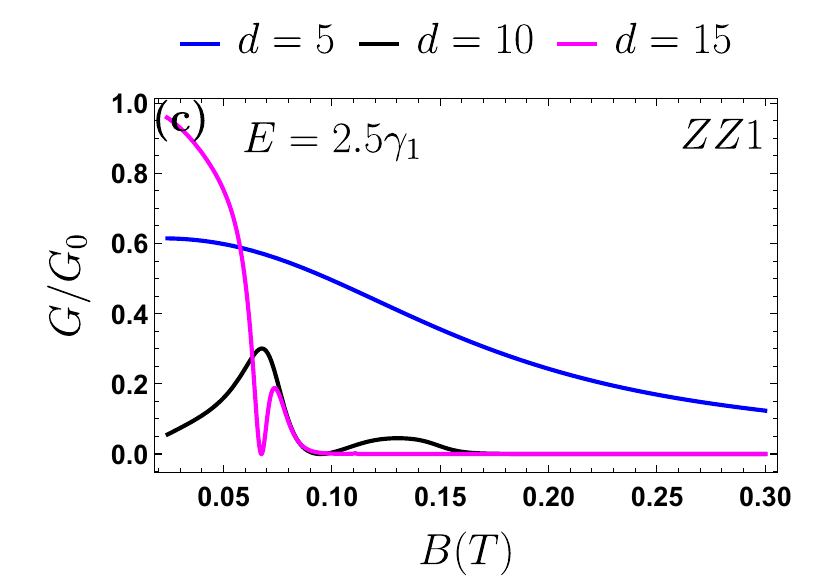}\includegraphics[width=0.5\linewidth]{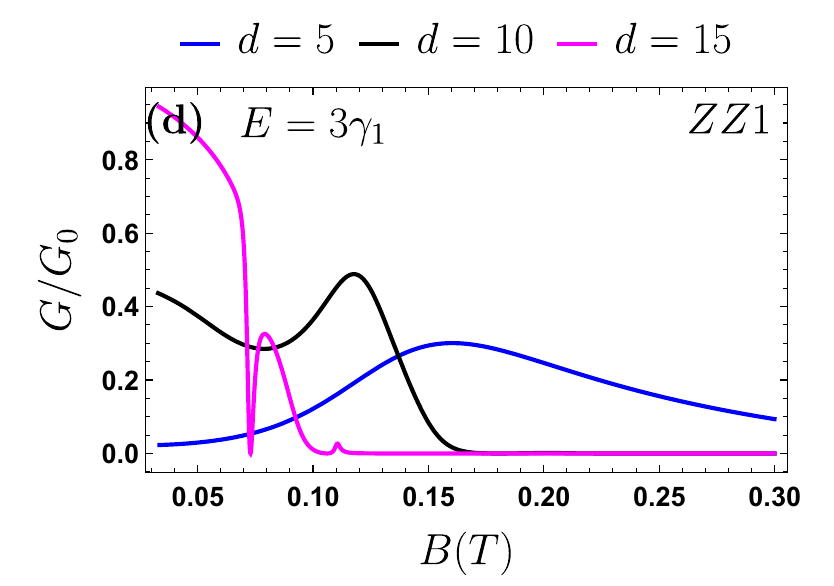}	
	\caption{(Color online) Conductance as a function of the magnetic field for ZZ1  for various width of BLG: $d=5$ nm (blue line), $d=10$ nm (black line), and $d=15$ nm (magenta line), and for various values of the Fermi energy, (a) $E=\gamma_1$, (b) $E=2\gamma_1$, (c) $E=2.5\gamma_1$, (d) $E=3\gamma_1$.}\label{G as a function of B}
\end{figure}
\begin{figure}[htp]
	\centering
	\includegraphics[width=0.5\linewidth]{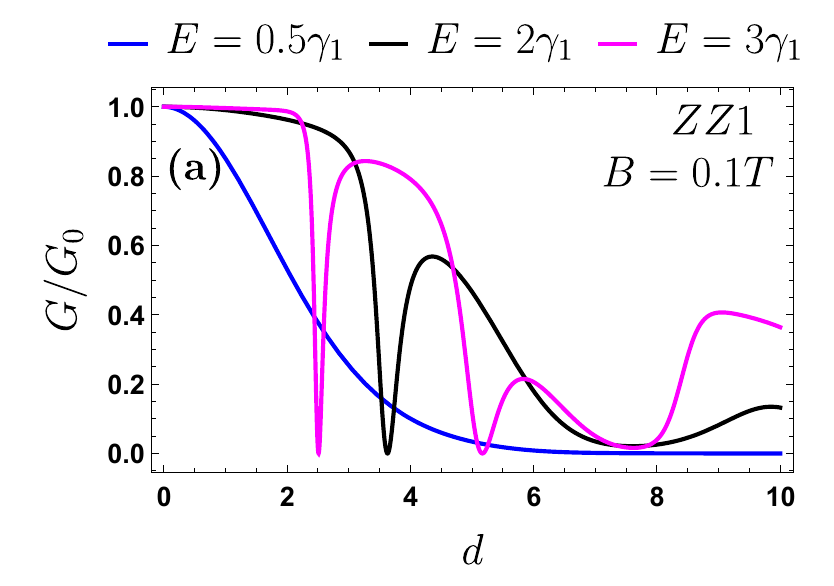}\includegraphics[width=0.5\linewidth]{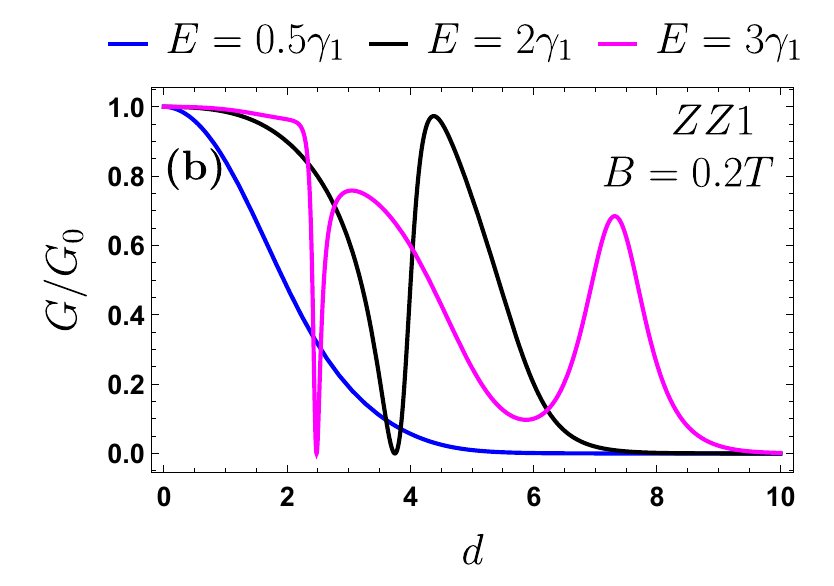}\\
	\includegraphics[width=0.5\linewidth]{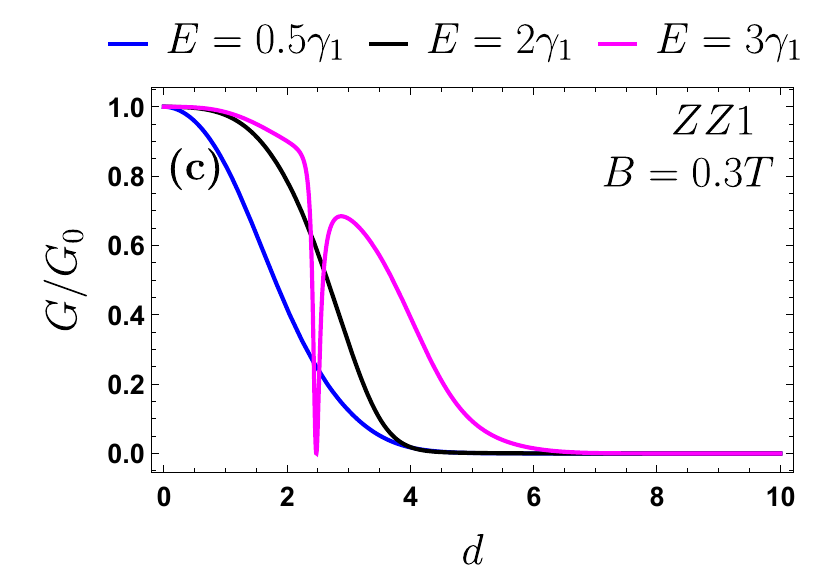}\includegraphics[width=0.5\linewidth]{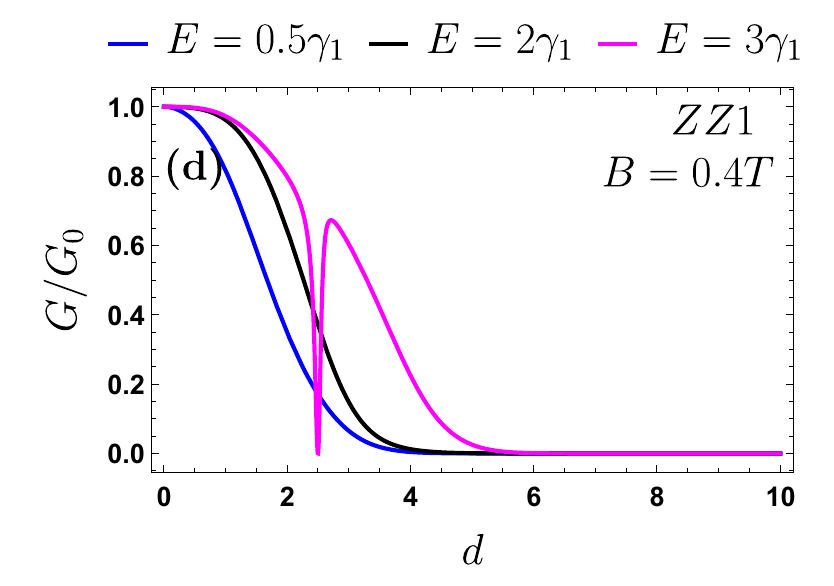}
	\caption{(Color online) Conductance as a function of the width of BLG $d$ for ZZ1 for various  Fermi energy values: $E=0.5\gamma_1$ nm (blue line), $E=2\gamma_1$ nm (black line), and $E=3\gamma_1$ nm (magenta line), and for various values the magnetic field $B$, (a) $B=0.1$ T, (b) $B=0.2$ T, (c) $B=0.3$ T, (d) $B=0.4$ T.}\label{G as a function of d}
\end{figure}
\begin{figure}[htp]
	\centering
	\includegraphics[width=0.5\linewidth]{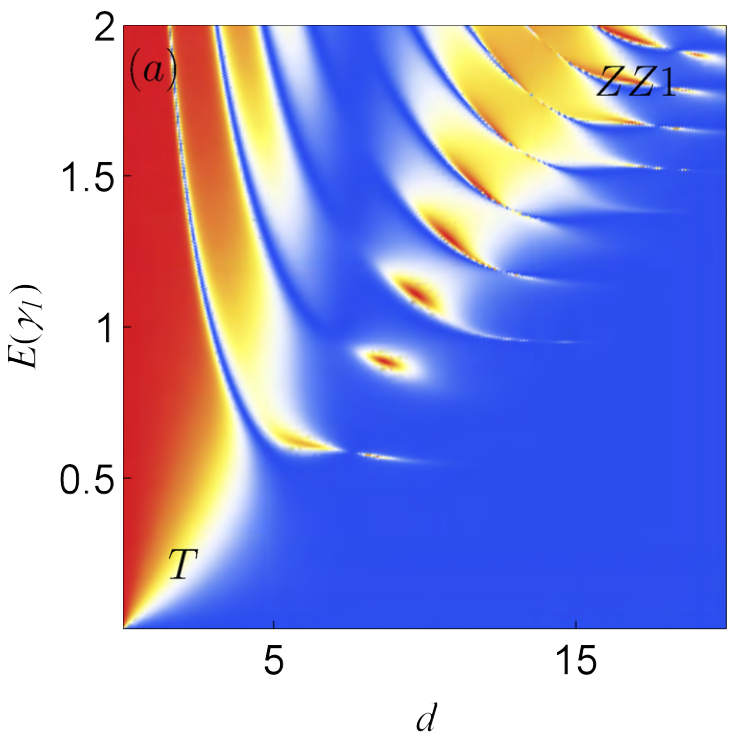}\includegraphics[width=0.61\linewidth]{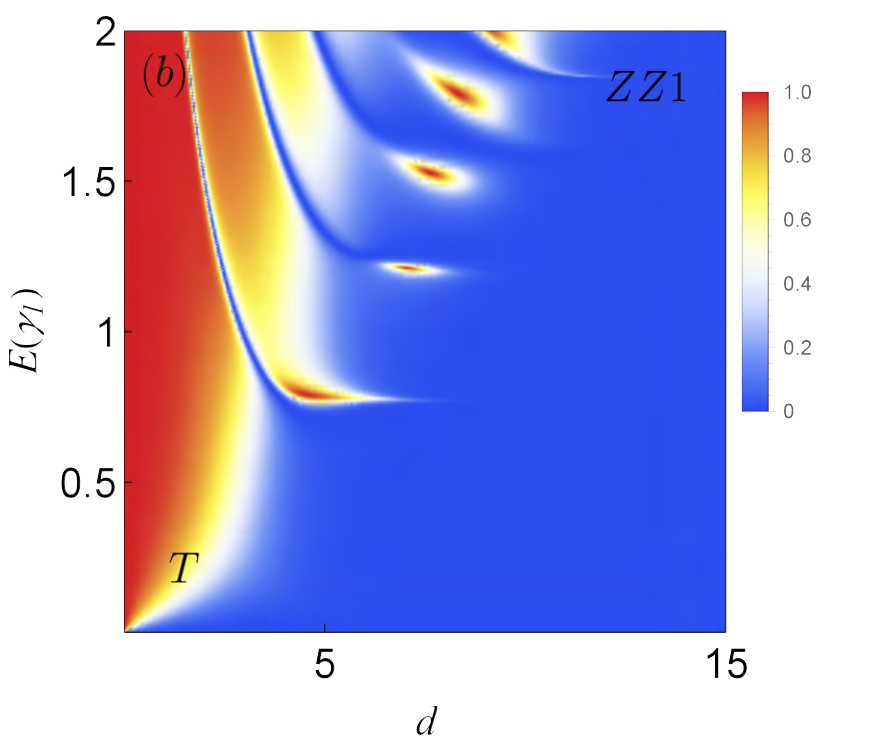}\\
	\includegraphics[width=0.5\linewidth]{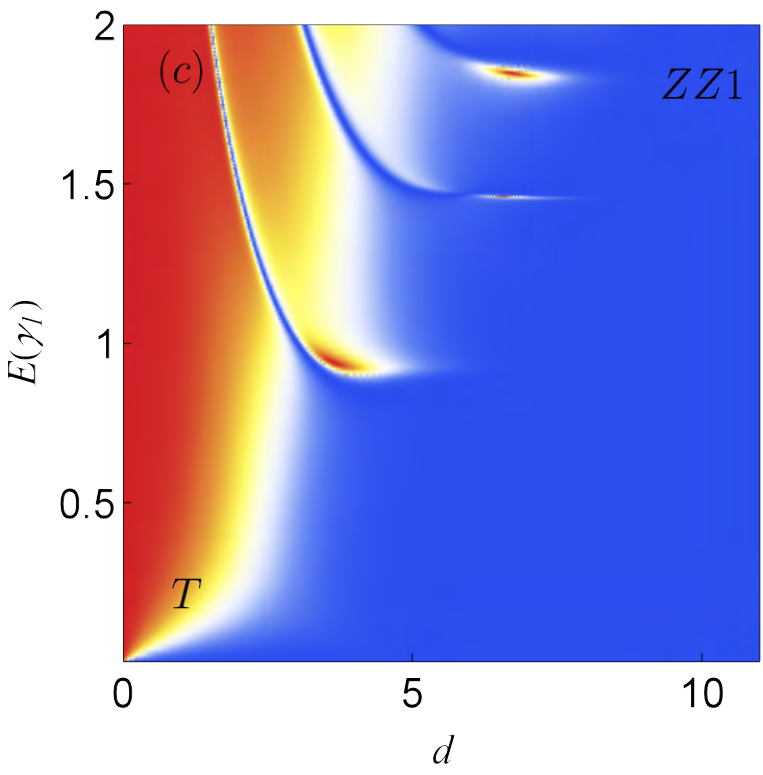}\includegraphics[width=0.5\linewidth]{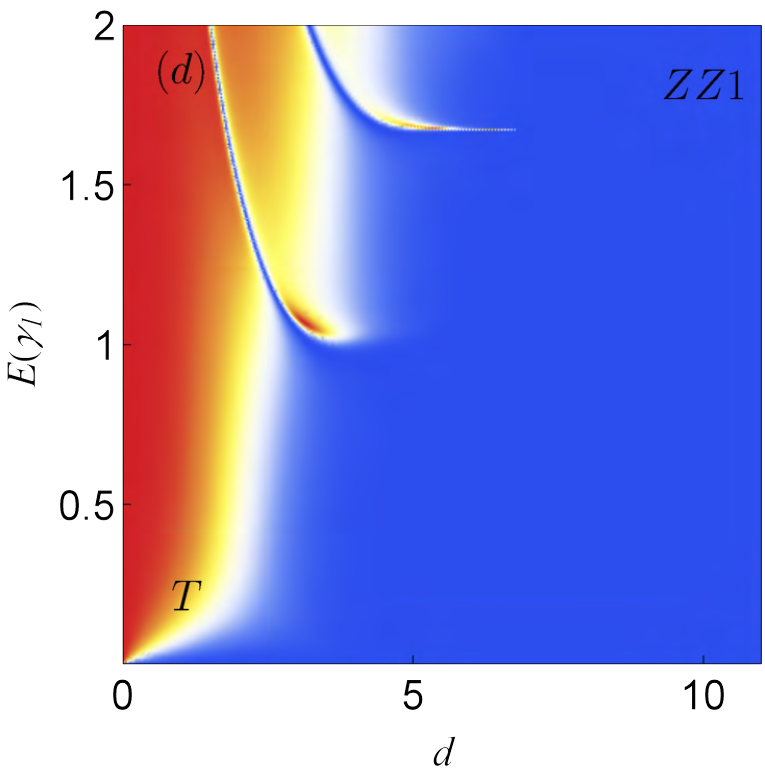}
	\caption{(Color online) Density plot of the transmission
		probability as a function of bilayer region length $d$, and
		Fermi energy $E$  for ZZ1,  (a) $B=0.1$ T,  (b) $B=0.2$ T,  (c) $B=0.3$ T,  (d) $B=0.4$ T.
		}\label{DPTZZ1}
\end{figure}
\begin{figure}[htp]
	\centering
	\includegraphics[width=0.5\linewidth]{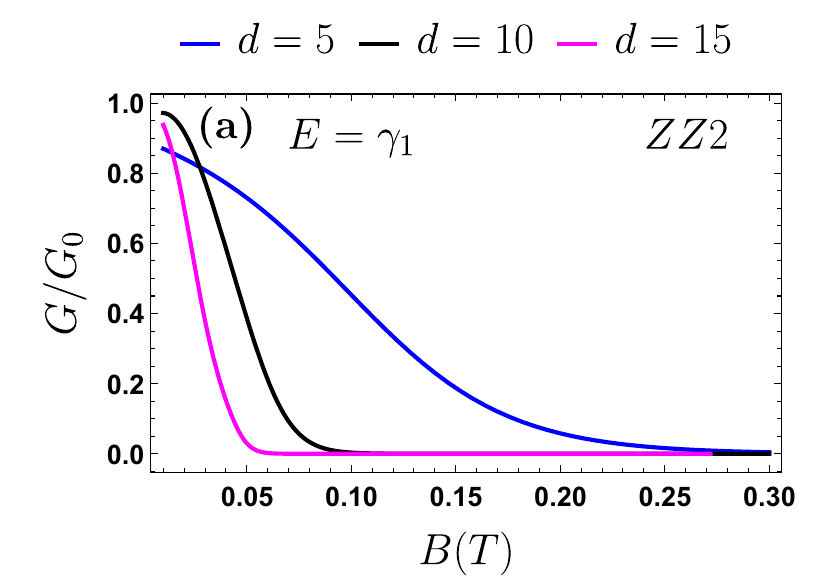}\includegraphics[width=0.5\linewidth]{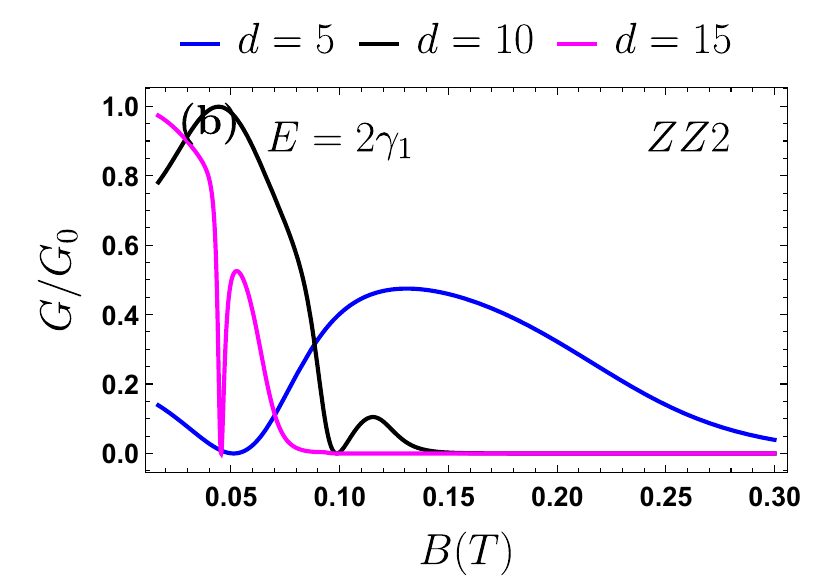}\\
	\includegraphics[width=0.5\linewidth]{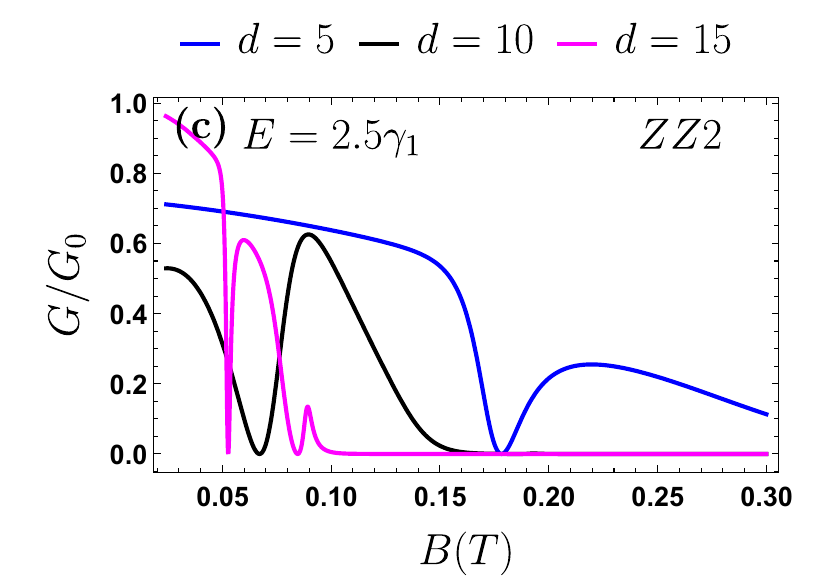}\includegraphics[width=0.5\linewidth]{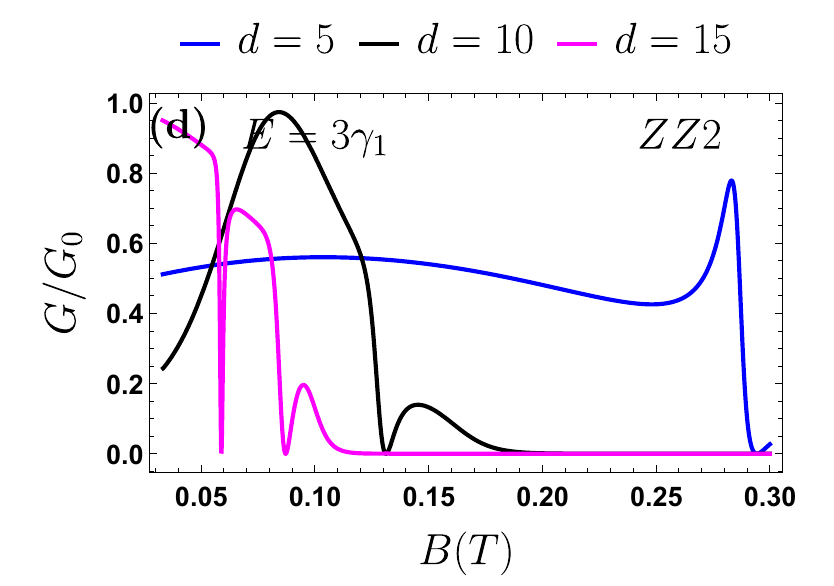}	
	\caption{(Color online) Conductance as a function of the magnetic field for ZZ2  for various width of BLG $d$ values: $d=5$ nm (blue  line), $d=10$ nm ( black line), and $d=15$ nm (magenta line), and for various values of the Fermi energy, (a) $E=\gamma_1$, (b) $E=2\gamma_1$, (c) $E=2.5\gamma_1$, (d) $E=3\gamma_1$.}\label{G as a function of B2}
\end{figure}
\begin{figure}[htp]
	\centering
	\includegraphics[width=0.5\linewidth]{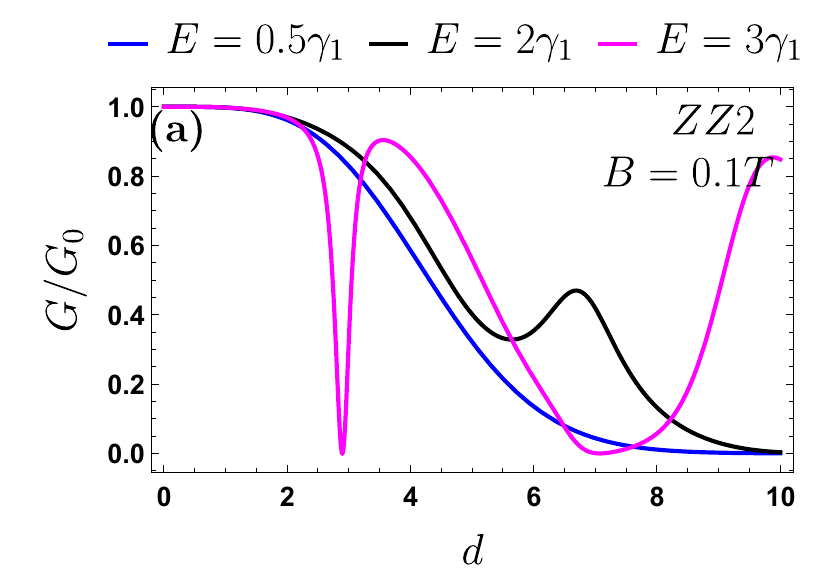}\includegraphics[width=0.5\linewidth]{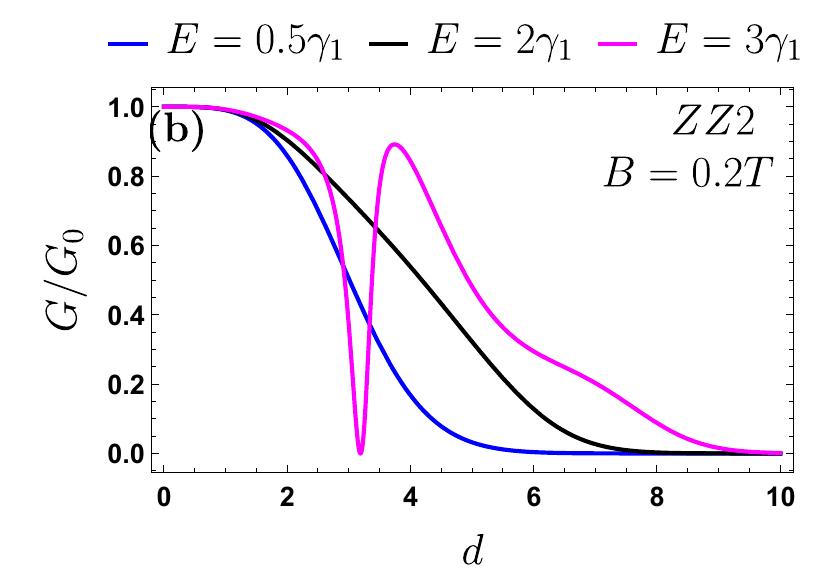}\\
	\includegraphics[width=0.5\linewidth]{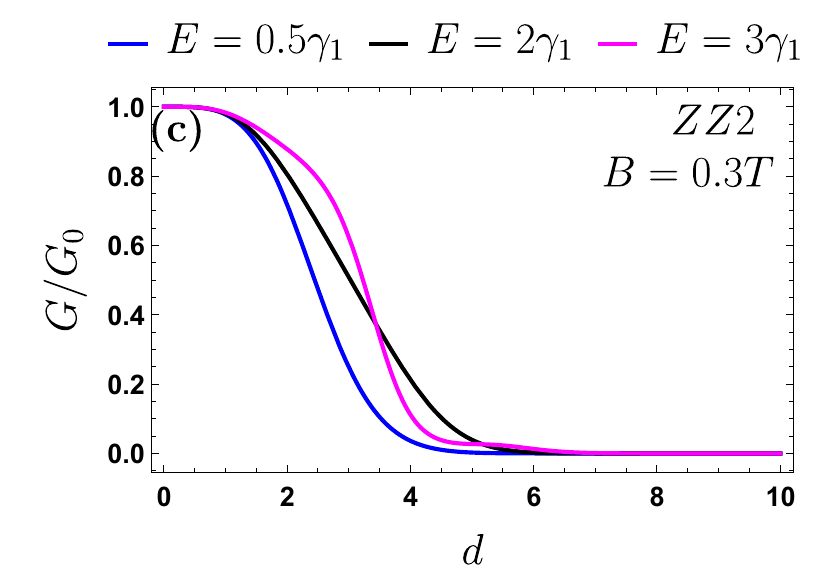}\includegraphics[width=0.5\linewidth]{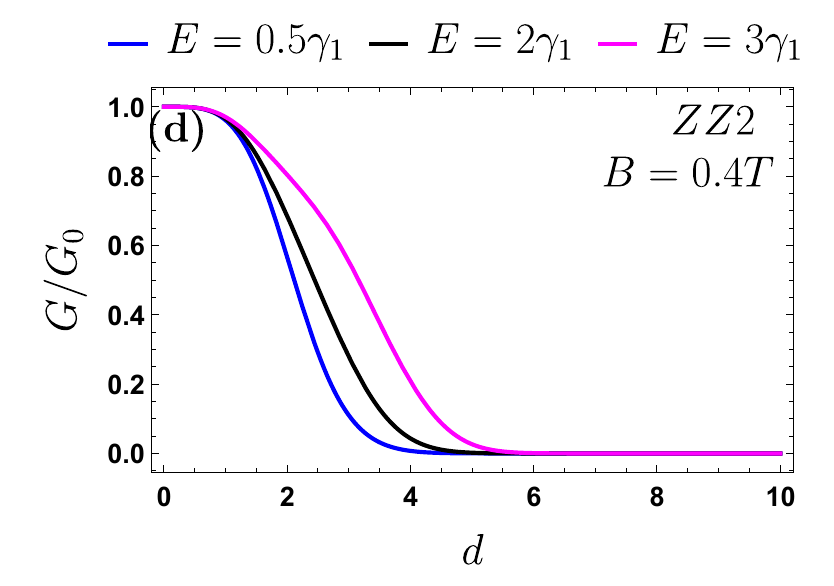}
	\caption{(Color online) Conductance as a function of the width of BLG $d$ for ZZ2 for various  Fermi energy values: $E=0.5\gamma_1$ nm (blue line), $E=2\gamma_1$ nm ( black line), and $E=3\gamma_1$ nm (magenta  line), and for various values the magnetic field $B$, (a) $B=0.1$ T, (b) $B=0.2$ T, (c) $B=0.3$ T, (d) $B=0.4$ T.}\label{G as a function of d2}
\end{figure}
\begin{figure}[htp]
	\centering
	\includegraphics[width=0.5\linewidth]{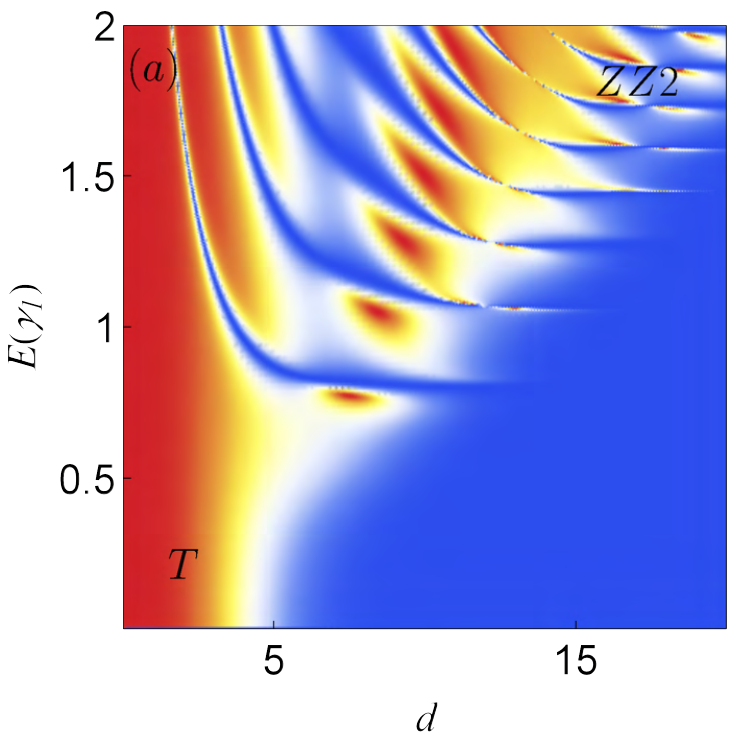}\includegraphics[width=0.61\linewidth]{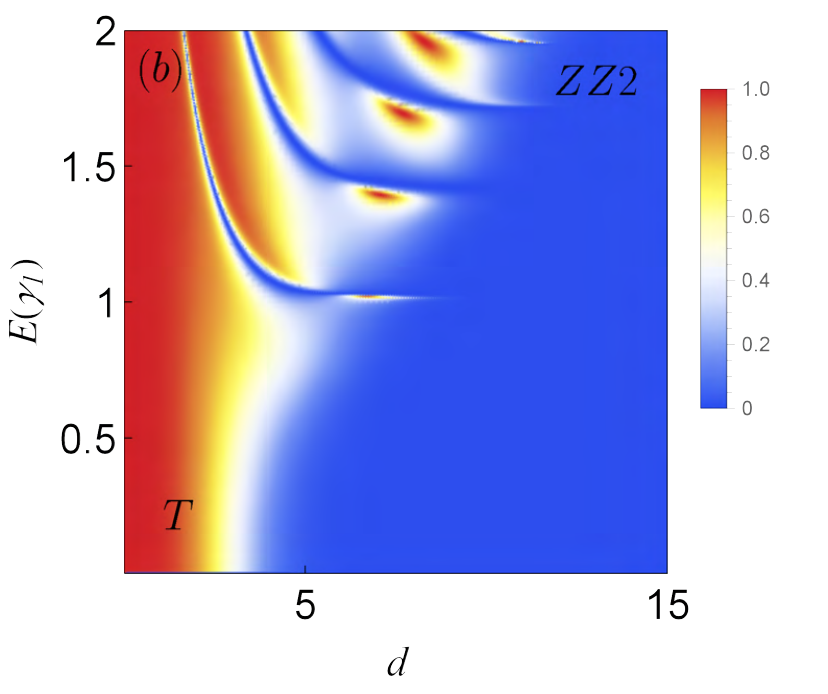}\\
	\includegraphics[width=0.5\linewidth]{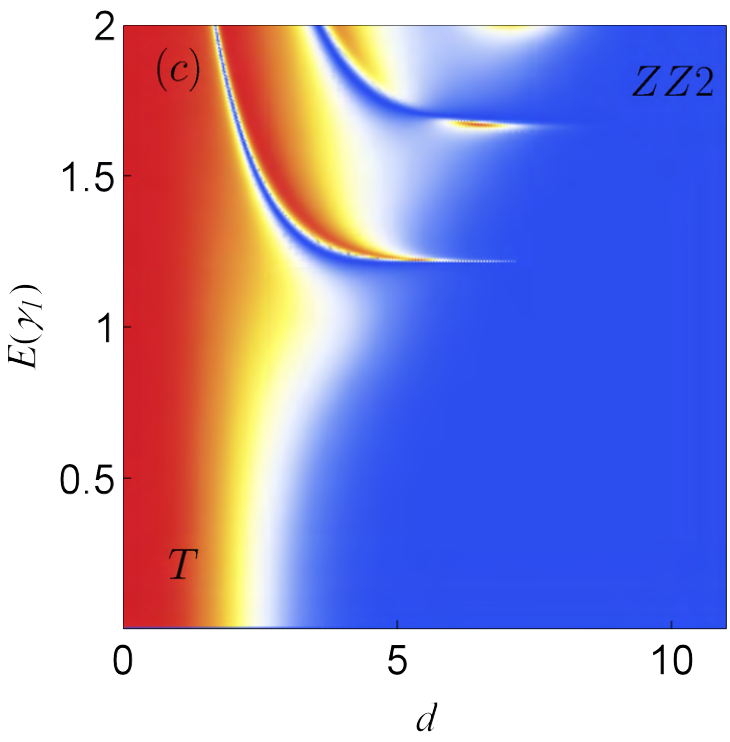}\includegraphics[width=0.5\linewidth]{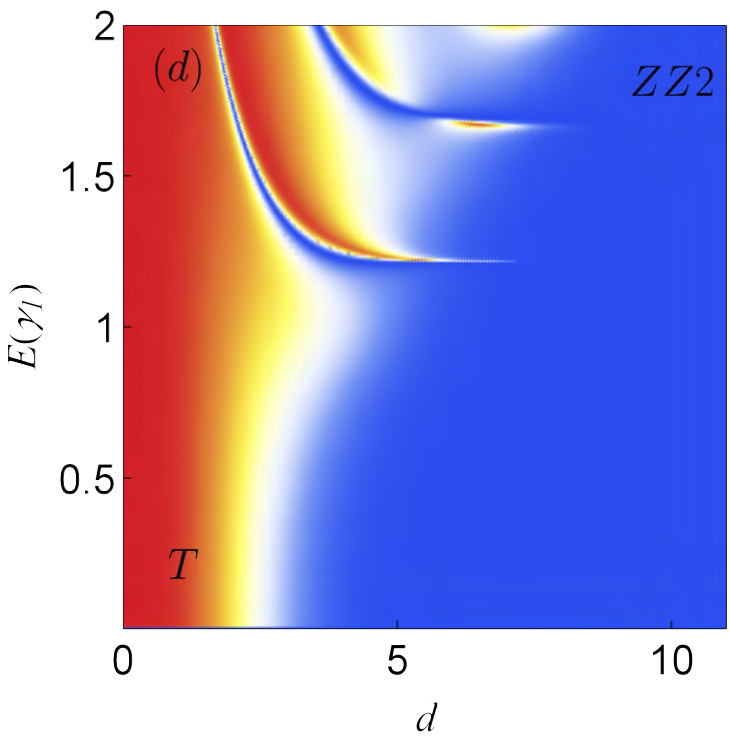}
	\caption{(Color online) Density plot of the transmission probability as a function of bilayer region length $d$, and
		Fermi energy $E$  for ZZ2, with (a) $B=0.1$ T,  (b) $B=0.2$ T,  (c) $B=0.3$ T,  (d) $B=0.4$ T.}\label{DPTZZ2}
\end{figure}
\begin{figure}[htp]
	\centering
	\includegraphics[width=0.5\linewidth]{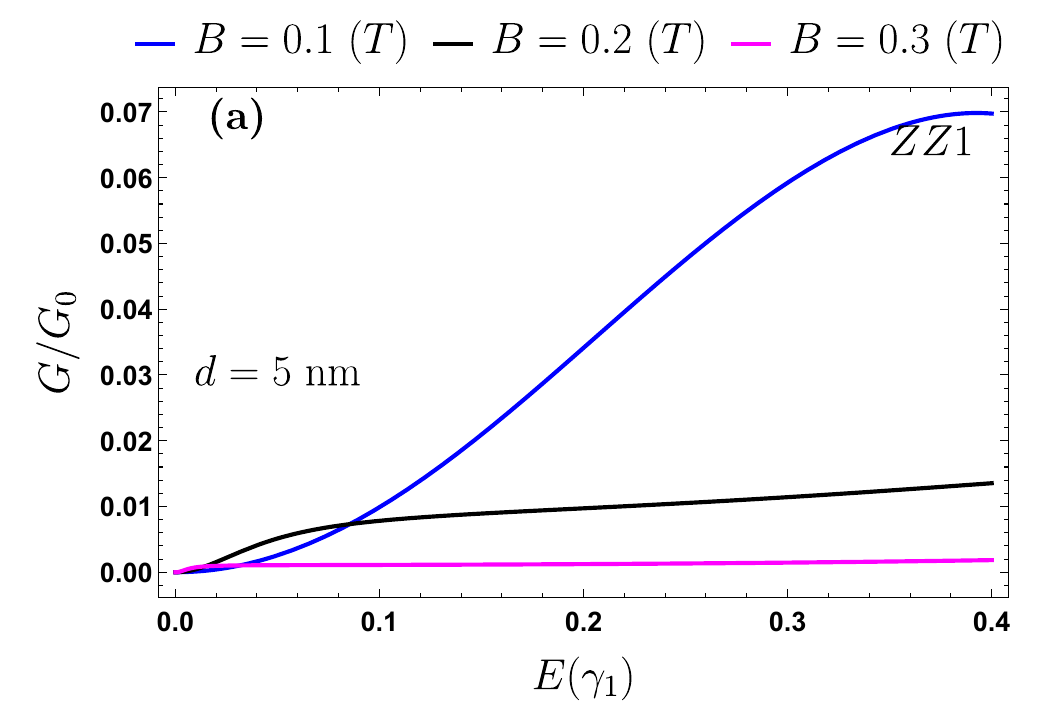}\includegraphics[width=0.5\linewidth]{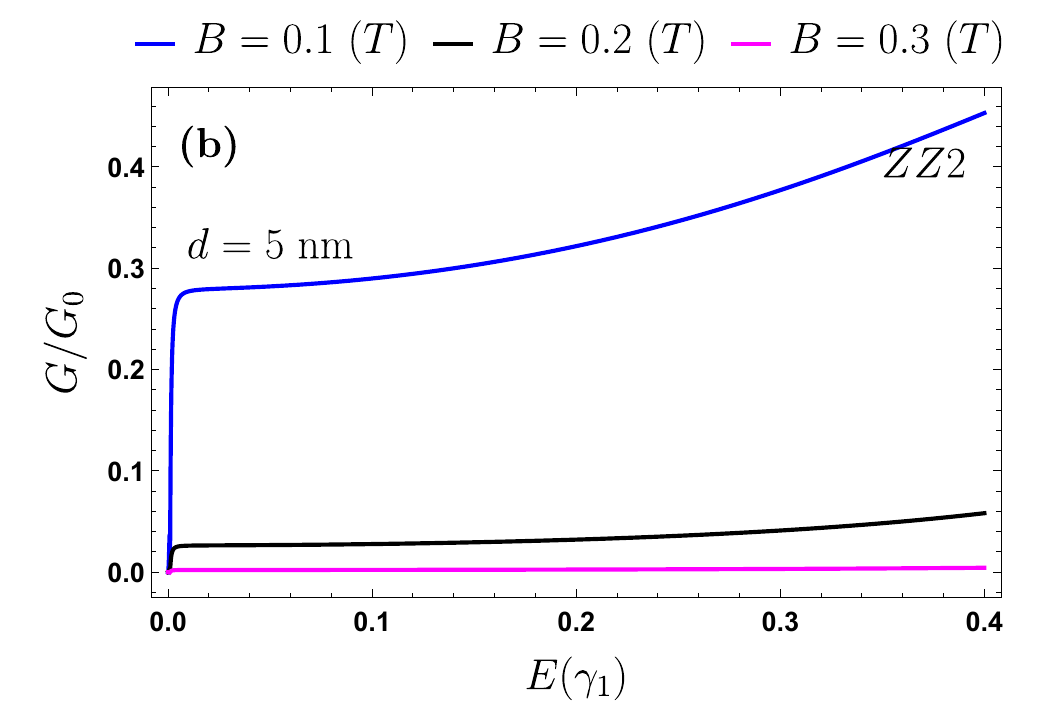}
	\caption{(Color online) Conductance as a function of the Fermi energy  for (a) ZZ1 and (b) ZZ2, for various values of the magnetic field :
		$B=0.1$ T (blue line), $B=0.2$ T (black line), and $B=0.3$ T
		(magenta line).}\label{TEZZI}
\end{figure}

\section{RESULTS AND DISCUSSION}\label{RESULTS AND DISCUSSION}
\subsection{Zigzag boundary, ZZ1, and ZZ2}
In this section, we analyze our key findings numerically and discuss them. In Fig. \ref{G as a function of B}, we show the conductance as a function of the magnetic field for different widths of the BLG varying the values of the Fermi energy $E$. {After careful analysis, noticeable variations in the conductance $G(E)$ at different energy values have become evident.}
 For example, the blue curves exhibit noticeable variations in all panels (energies) of Fig. \ref{G as a function of B}, indicating a clear dependence of the conductance on energy. 
This is a manifestation of Klein tunneling \cite{Ben 1, Ben 2}. In Fig. \ref{G as a function of B} (a), for $E=\gamma_1$, the result shows $0$ conductance and there are no antiresonances because there is only one propagating channel at the BLG. However, it is not the same case for the $E>\gamma_1$. Fig. \ref{G as a function of B} (b), (c), and (d), the $G(E)$ presents antiresonances that appear with zero conductance for large values of the magnetic field, because of the coexistence of two propagating channels. Fig. \ref{G as a function of d} shows the conductance as a function of the  BLG width for the ZZ1 boundary for four different values of the magnetic field. We notice that for $E=0.5\gamma_1$, for which there is only one propagating channel in the BLG, the conductance shows an exponential decay of the resonance for sufficiently long widths and all magnetic field values, as shown in Figs. \ref{G as a function of d} (a) and (b). However, for the highest energies, the resonances become important and dependent on the magnetic field. We observe that the shapes and the numbers of the resonances change from $B=0.1$ T to $B=0.4$ T. For sufficiently long widths, i.e.,  $d>4$ nm the conductance $G$ tends to $0$. This result shows agreement with our previous results \cite{N. Benlakhouy1}. The presence of the magnetic field in the bilayer graphene in this SLG-BLG-SLG affects the conductance and removes the periods of the antiresonances as seen in \cite{J. W. Gonzalez}.

We exhibit in Fig. \ref{DPTZZ1} the density plot of the transmission probability as a function of the Fermi energy $E$ and the width of the BLG 
to examine the impact of the magnetic field. As displayed in the plots  there are two separate regions of energy, set by the interlayer coupling $E<\gamma_1$, there are no antiresonances due to the existence of just one propagation channel at the BLG, in contrast to  $E>\gamma_1$, where the coexistence of two propagating eigenchannels in the BLG lets the zero antiresonances appear. The behavior is comparable to that investigated by González {\it et al} \cite{J. W. Gonzalez}, with a clear distinction between the spatial periods, due to the presence of the magnetic field, and also we observe that transmission is completely suppressed for large widths of the BLG ($d>2.5$) and {the Klein tunneling diminishes or becomes less prominent}.
 Hence, we conclude that in the presence of the magnetic field, Klein tunneling is hampered, and instead Febry-Pérot resonances \cite{I. Snyman} appear for $E>\gamma_1$. {The phenomenon of interest has been studied in detail in a recent work \cite{A. Pena}, where the authors demonstrated remarkable progress in understanding and controlling tunneling behavior in a similar system}.

We will  turn our attention now to the zigzag boundary ZZ2, for which we plot
the conductance as a function of the magnetic field in Fig. \ref{G as a function of B2} for three different widths of the BLG, taking into account four different values of the Fermi energy. In Fig. \ref{G as a function of B2} (a) the conductance shows maxima for $B<0.10$ T, in contrast to the ZZ1 boundary, by increasing the width of BLG and the Fermi energy, the conductance shows oscillations in the ZZ2 case. These results show that the transmission
probability depends more strongly on boundaries, and  confinement is more important in the ZZ1 boundary than the ZZ2 boundary, which is in agreement with \cite{Nakanishi} results. In Fig. \ref{G as a function of d2},  we plot the conductance as a function of the width $d$  using the same parameters as in Fig. \ref{G as a function of d}. We observe fewer resonances than in the ZZ1 case for $B\leqslant 0.3$ T. By increasing the BLG width,  we see that our conductance also
vanishes for the ZZ2 boundary.

Fig. \ref{DPTZZ2} displays the density plots for the ZZ2 case. There are two important differences with respect to the ZZ1 case, see Fig. \ref{DPTZZ1}. First, for the energy region $E<\gamma_1$ the transmission is more important, with an obvious difference in the contrast of the spatial resonances. We note too that the transmission with respect to $d$ is more similar to the $ZZ1$ case, but both results indicate strong confinement at the ZZ1 boundary.
{In Fig. \ref{TEZZI}, we present the conductance as a function of the Fermi energy $E$. In the  analysis conducted for ZZ1 and ZZ2, intriguing trends emerge. Notably, the blue line, corresponding to $B=0.1$ T, exhibits a prominent peak, suggesting an enhanced conductance in ZZ1 compared to ZZ2. On the other hand, for both ZZ1 and ZZ2, the black and magenta lines representing $B=0.2$~T and $B=0.3$~T, respectively, show a clear convergence towards zero with increasing Fermi energy. These findings indicate that higher magnetic fields lead to a rapid decrease in conductance for both ZZ1 and ZZ2. This suggests a significant influence of the magnetic field strength on the conductance behavior in both regions, leading to distinctly different conductive properties between ZZ1 and ZZ2.}
\subsection{Combined Results}
\begin{figure}[htp]
	\centering
	\includegraphics[width=0.5\linewidth]{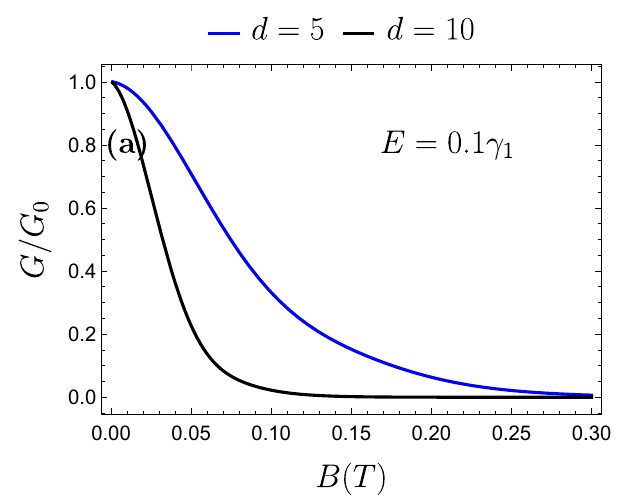}\includegraphics[width=0.5\linewidth]{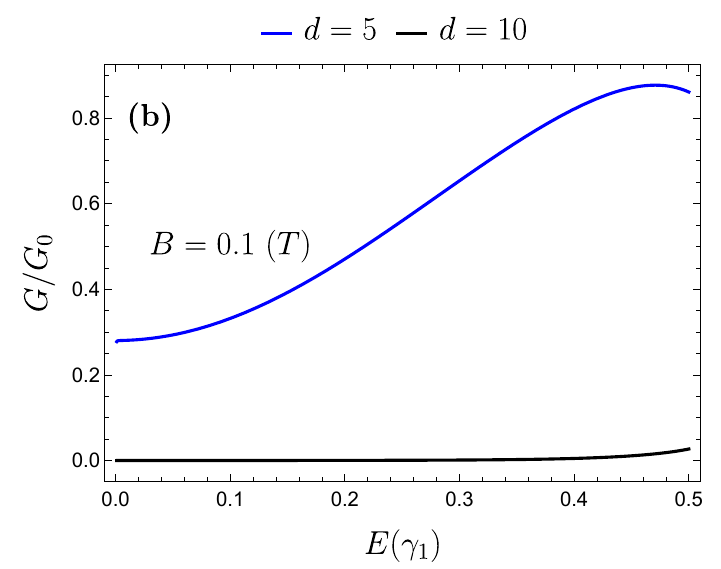}
	\caption{(Color online) (a) Conductance as a function of the magnetic field for
		$d=5$ nm (blue line), $d=10$ nm (black line), with $E=0.1\gamma_1$. (b) Conductance as a function of the Fermi energy for
		$d=5$ nm (blue line), $d=10$ nm (black line), with $B=0.1$ (T).}\label{ZZab}
\end{figure}
{In this subsection, we consolidate our analysis by considering the interfaces ZZ1 and ZZ2 within the BLG system. We aim to elucidate the behavior of electron transmission through the BLG positioned between two SLG regions under the influence of a perpendicular magnetic field. The combined findings shed light on the overall transport properties of the system.
	Fig. \ref{ZZab} (a) showcases the total conductance as a function of the magnetic field for specific separations, $d=5$ nm (blue line) and $d=10$ nm (black line), both at $E=0.1\gamma_1$. These results unveil intriguing trends in electron transport behavior within the BLG. The observed conductance variation with the magnetic field exhibits an exponential-like trend, signifying a pronounced sensitivity to the applied magnetic field.
	Fig. \ref{ZZab} (b) presents the conductance concerning the Fermi energy for $d=5$ nm and $d=10$ nm, maintaining a constant magnetic field of $B=0.1$ T. These findings offer valuable insights into distinct electron transport behaviors observed within the BLG system associated with varying widths.
	Specifically, for the $d=10$ nm case, an initial conductance of zero denotes a lack of electron transmission at the lowest Fermi energy considered. This observation strongly suggests the potential formation of a bandgap in the bilayer system at this particular width.
	Conversely, for the $d=5$ nm case, a contrasting trend emerges. The conductance exhibits an initial sharp increase as the Fermi energy rises, indicating the presence of available electron states conducive to electron transport within the bilayer system at this specific width.}
\section{Conclusion}\label{Conclusion}
{After closely analyzing the impact of a perpendicular magnetic field within BLG surrounded by SLG regions, our investigation revealed notable insights. Our focus on this configuration highlighted the specific influence of the magnetic field solely within the BLG amidst the surrounding SLG regions. This examination sheds light on the magnetic field's selective impact within the BLG context, contributing to a deeper understanding of its effect in this composite structure}. We looked at both types of zigzag boundaries. Starting with ZZ1, we showed that the conductance as a function of the magnetic field at normal incidence $(k_y = 0)$ seems dependent on energy. Due to pseudospin conservation, this is an instance of Klein tunneling. For $E=\gamma_1$, the result shows zero conductance and there are no antiresonances because there is only one propagating channel at the BLG, but for $E>\gamma_1$, $G(E)$ presents antiresonances that appear with zero conductance for large values of the magnetic field because of the coexistence of two propagating channels. 

 As a function of the  BLG width, we have found that for $E=0.5\gamma_1$, for which there is only one propagating channel in the BLG, the conductance shows an exponential decay of the resonance for a sufficiently long width and all magnetic field values. For the highest energies, the resonances become important and dependent on the magnetic field. Our results also show that the shapes and numbers of the resonances change from $B=0.1$ T to $B=0.4$ T. For sufficiently long widths, i.e.,  $d>4$ nm, the conductance $G$ tends to $0$.  For the zigzag boundary ZZ2, we found distinct behaviors compared with the ZZ1 boundary. We observed maxima for small values of the magnetic field in the conductance plot, in contrast to the ZZ1 boundary, and by increasing the width of BLG and the Fermi energy, the conductance shows oscillatory behavior in the ZZ2 feature. Our results showed that the transmission probability depends more strongly on boundaries, and confinement is more important in the ZZ1 boundary than in the ZZ2 boundary. We have also analyzed the conductance as a function of the width and length $d$. We have observed fewer resonances than in the ZZ1 case for $B\leqslant 0.3$ T. Increasing the BLG width, our conductance $G(E)$ vanishes also for the ZZ2 boundary.
 
 {Our analysis of the conductance behavior for ZZ1 and ZZ2, varying with the Fermi energy, revealed interesting trends. Notably, ZZ1 demonstrated higher conductance compared to ZZ2, while both regions showed a rapid decrease in conductance with increasing Fermi energy. These results demonstrate the important influence of the magnetic field on the transport properties of ZZ1 and ZZ2, leading to different conductances between the two locations.}

\appendix\label{Appendix}

\section{Transfer matrix for SLG and BLG junction}
In this Appendix, we briefly review the main steps of
our analytical calculations. In order to determine the transmission  probability, we impose the appropriate boundary conditions in the context of the transfer matrix approach. More explicitly, for boundary ZZ1 Eqs. \ref{ZZ1x0} and \ref{ZZ1xd}, we  obtain six equations with six unknowns, at $x=0$
\begin{widetext}
\begin{align}
\label{ZZ11}&\alpha^{-}-r \alpha^{+}=c_{+} \nu \lambda_{+} D_{\lambda_{+}-1}(Z_1)+c_{-} \nu^* \lambda_{+} D_{\lambda_{+}-1}(-Z_1)+d_{+} \nu\lambda_{-}   D_{\lambda_{-}-1}(Z_1)+d_{-}  \nu^* \lambda_{-} D_{\lambda_{-}-1}(-Z_1),  \\
\label{ZZ12}&      1+r=c_{+} D_{\lambda_{+}}(Z_1)+c_{-} D_{\lambda_{+}}(-Z_1) +d_{+} D_{\lambda_{-}}(Z_1)+d_{-}D_{\lambda_{-}}(-Z_1),\\
\label{ZZ13}&	 	0=c_{+} \nu^{*} \zeta^{+}  D_{\lambda_{+}+1}(Z_1)+c_{-} \nu^{*}\zeta^{+}  D_{\lambda_{+}+1}(-Z_1)+d_{+} \nu \zeta^{-} D_{\lambda_{-}+1}(Z_1)+d_{-} \nu \zeta^{-}\lambda_{-}  D_{\lambda_{-}+1}(-Z_1),
\end{align}
and for $x=d$  
\begin{align}
\label{ZZ14}&	c_{+} \nu \lambda_{+} D_{\lambda_{+}-1}(Z_2)+c_{-} \nu^* \lambda_{+} D_{\lambda_{+}-1}(-Z_2)+d_{+} \nu\lambda_{-}   D_{\lambda_{-}-1}(Z_2)+d_{-}  \nu^* \lambda_{-} D_{\lambda_{-}-1}(-Z_2) =t\alpha^{-}e^{ik_{x}d},  \\
\label{ZZ15}&  c_{+} D_{\lambda_{+}}(Z_2)+c_{-} D_{\lambda_{+}}(-Z_2) +d_{+} D_{\lambda_{-}}(Z_2)+d_{-}D_{\lambda_{-}}(-Z_2)=te^{ik_{x}d},\\
\label{ZZ16}&  c_{+} \nu^{*} \zeta^{+}  D_{\lambda_{+}+1}(Z_2)+c_{-} \nu^{*}\zeta^{+}  D_{\lambda_{+}+1}(-Z_2)+d_{+} \nu \zeta^{-} D_{\lambda_{-}+1}(Z_2)+d_{-} \nu \zeta^{-}\lambda_{-}  D_{\lambda_{-}+1}(-Z_2)=0. 
\end{align}
\end{widetext}
We note that the eigenspinors for SLG at $(x=0)$ consist of the incident and reflected plane waves, then the Eqs. in \ref{phiA-PhiB} rewritten as
\begin{align}
&\phi_A(x=0)=\alpha^{-}-r\alpha^{+},\\
&\phi_B(x=0)=1+r,	
\end{align}
and consist of the transmitted waves at $(x=d)$
\begin{align}
&\phi_A(x=d)=t\alpha^{-}e^{ik_xd},\\
&\phi_B(x=d)=te^{ik_xd},	
\end{align}
with $r$ and $t$ denote the reflection and transmission coefficients. Eqs. \ref{ZZ11} and \ref{ZZ12} as well as Eqs. \ref{ZZ14} and \ref{ZZ15} can be reformulated using the transfer matrix approach  \cite{Yu Wang}
\begin{align}
\label{MT1}&\mathcal{G}^{\text{SLG}} \mathcal{P}_{x=0}\left[\begin{array}{l}
1 \\
r
\end{array}\right]=\mathcal{G}_{+,Z_1}^{\text{BLG}}\left[\begin{array}{l}
c_{+} \\
c_{-}
\end{array}\right]+\mathcal{G}_{-,Z_1}^{\text{BLG}}\left[\begin{array}{l}
d_{+} \\
d_{-}
\end{array}\right], \\
\label{MT2}&\mathcal{G}_{+,Z_2}^{\text{BLG}}\left[\begin{array}{l}
c_{+} \\
c_{-}
\end{array}\right]+\mathcal{G}_{-,Z_2}^{\text{BLG}}\left[\begin{array}{l}
d_{+} \\
d_{-}
\end{array}\right]=\mathcal{G}^{\text{SLG}} \mathcal{P}_{x=d}\left[\begin{array}{l}
t \\
0
\end{array}\right],
\end{align}
with
\begin{align}
&\mathcal{G}^{\text{SLG}}=\begin{pmatrix}
\alpha^{-} & -\alpha^{+}  \\
1 & 1   \\
\end{pmatrix},
\\
&
\mathcal{P}_{x}=\begin{pmatrix}
e^{ik x}  & 0 \\
0 & e^{-ik x}    \\
\end{pmatrix},
\\
&
\mathcal{G}_{\pm, Z_{1/2}}^{\text{BLG}}=\begin{pmatrix}
\nu\lambda_{\pm}D_{\lambda_{\pm}-1}(Z_{1/2})  & \nu^{*}\lambda_{\pm}D_{\lambda_{\pm}-1}(-Z_{1/2}) \\
D_{\lambda_{\pm}-1}(Z_{1/2}) &D_{\lambda_{\pm}-1}(-Z_{1/2}) \\
\end{pmatrix}.
\end{align}
Combining Eqs. \ref{ZZ13} and \ref{ZZ16}, its matrix counterpart can be expressed as
\begin{equation}
\left[\begin{array}{l}
d_{+} \\
d_{-}
\end{array}\right]=\mathcal{N}_{\text{ZZ1}}\left[\begin{array}{l}
c_{+} \\
c_{-}
\end{array}\right],\label{MT3}
\end{equation}
with
\begin{equation}
\begin{aligned}
\mathcal{N}_{\text{ZZ1}}= & -\left[\begin{pmatrix}
\nu \zeta^{-} D_{\lambda_{-}+1}(Z_1) & \nu \zeta^{-} D_{\lambda_{-}+1}(-Z_{1}) \\
\nu \zeta^{-} D_{\lambda_{-}+1}(Z_2) & \nu \zeta^{-} D_{\lambda_{-}+1}(Z_2)   \\
\end{pmatrix}\right]^{-1} \\
& \times\left[\begin{pmatrix}
\nu^{*} \zeta^{+} D_{\lambda_{+}+1}(Z_1)  & \nu^{*} \zeta^{+} D_{\lambda_{+}+1}(-Z_1) \\
\nu^{*} \zeta^{+} D_{\lambda_{+}+1}(Z_2) & \nu^{*} \zeta^{+} D_{\lambda_{+}+1}(-Z_2)    \\
\end{pmatrix}\right].
\end{aligned}
\end{equation}
Combining Eqs. \ref{MT1}, \ref{MT2}, and \ref{MT3}, the transfer matrix of the our structure can be obtained as
\begin{equation}
\left[\begin{array}{c}
1 \\
r
\end{array}\right]=\mathcal{M}_{\text{ZZ1}}\left[\begin{array}{c}
t \\
0
\end{array}\right],\label{MZZ1}
\end{equation}
with
\begin{align}
\mathcal{M}_{\text{ZZ1}}= & \left(\mathcal{G}^{\text{SLG}} \mathcal{P}_{x=0}\right)^{-1} \left[\mathcal{G}_{+, Z_1}^{\text{BLG}}+\mathcal{G}_{-, Z_1}^{\text{BLG}} \mathcal{N}_{\text{ZZ1}}\right]\\
& \times\left[\mathcal{G}_{+, Z_2}^{\text{BLG}}+\mathcal{G}_{-, Z_2}^{\text{BLG}} \mathcal{N}_{\text{ZZ1}}\right]^{-1} \mathcal{G}^{\text{SLG}}\mathcal{P}_{x=d}.
\end{align}
From Eq. \ref{MZZ1}, the transmission coefficient can be derived as
\begin{equation}\label{A18}
 t_{\text{ZZ1}} = \mathcal{M}^{-1}_{\text{ZZ1}}.
\end{equation}
In the same way, by requiring the continuity using Eqs. \ref{ZZ2x0} and \ref{ZZ2xd}, we get
\begin{widetext}
	\begin{align}
	&\alpha^{-}-r \alpha^{+}=c_{+} \nu \lambda_{+} D_{\lambda_{+}-1}(Z_1)+c_{-} \nu^* \lambda_{+} D_{\lambda_{+}-1}(-Z_1)+d_{+} \nu\lambda_{-}   D_{\lambda_{-}-1}(Z_1)+d_{-}  \nu^* \lambda_{-} D_{\lambda_{-}-1}(-Z_1),  \\
	&	1+r=c_{+} D_{\lambda_{+}}(Z_1)+c_{-} D_{\lambda_{+}}(-Z_1) +d_{+} D_{\lambda_{-}}(Z_1)+d_{-}D_{\lambda_{-}}(-Z_1),\\
	&0=c_{+} \zeta^{+} D_{\lambda_{+}}(Z_1)+c_{-} \zeta^{+}  D_{\lambda_{+}}(-Z_1)+d_{+} \zeta^{-}  D_{\lambda_{-}}(Z_1)+d_{-} \zeta^{-}  D_{\lambda_{-}}(-Z_1),
	\end{align}
	and for $x=d$  
	\begin{align}
	&	c_{+} \nu \lambda_{+} D_{\lambda_{+}-1}(Z_2)+c_{-} \nu^* \lambda_{+} D_{\lambda_{+}-1}(-Z_2)+d_{+} \nu\lambda_{-}   D_{\lambda_{-}-1}(Z_2)+d_{-}  \nu^* \lambda_{-} D_{\lambda_{-}-1}(-Z_2) =t\alpha^{-}e^{ik_{x}d},  \\
	&c_{+} D_{\lambda_{+}}(Z_2)+c_{-} D_{\lambda_{+}}(-Z_2) +d_{+} D_{\lambda_{-}}(Z_2)+d_{-}D_{\lambda_{-}}(-Z_2)=te^{ik_{x}d},\\
	&c_{+} \zeta^{+} D_{\lambda_{+}}(Z_2)+c_{-} \zeta^{+}  D_{\lambda_{+}}(-Z_2) +d_{+} \zeta^{-} D_{\lambda_{-}}(Z_2)+d_{-} \zeta^{-}  D_{\lambda_{-}}(-Z_2)=0. \\
	\end{align}
\end{widetext}
All these equations can be written in compact form by introducing the transfer matrix 
\begin{align}
\mathcal{M}_{\text{ZZ2}}= & \left(\mathcal{G}^{\text{SLG}} \mathcal{P}_{x=0}\right)^{-1} \left[\mathcal{G}_{+, Z_1}^{\text{BLG}}+\mathcal{G}_{-, Z_1}^{\text{BLG}} \mathcal{N}_{\text{ZZ2}}\right]\notag\\
& \times\left[\mathcal{G}_{+, Z_2}^{\text{BLG}}+\mathcal{G}_{-, Z_2}^{\text{BLG}} \mathcal{N}_{\text{ZZ2}}\right]^{-1} \mathcal{G}^{\text{SLG}}\mathcal{P}_{x=d},
\end{align}
where
\begin{equation}
\begin{aligned}
\mathcal{N}_{\text{ZZ2}}= & -\left[\begin{pmatrix}
\zeta^{-} D_{\lambda_{-}}(Z_1) &  \zeta^{-} D_{\lambda_{-}}(-Z_1) \\
\zeta^{-} D_{\lambda_{-}}(Z_2) &  \zeta^{-} D_{\lambda_{-}}(-Z_2)   \\
\end{pmatrix}\right]^{-1} \\
& \times\left[\begin{pmatrix}
\zeta^{+} D_{\lambda_{+}}(Z_1) &  \zeta^{+} D_{\lambda_{+}}(-Z_1) \\
\zeta^{+} D_{\lambda_{+}}(Z_2) &  \zeta^{+} D_{\lambda_{+}}(-Z_2)   \\
\end{pmatrix}\right].
\end{aligned}
\end{equation}
As a result, we can express  the transmission coefficient $t_{\text{ZZ2}}$ as 
\begin{equation}\label{A29}
t_{\text{ZZ2}} = \mathcal{M}^{-1}_{\text{ZZ2}}.
\end{equation}
\end{document}